\documentclass[aps,preprint,prd,showpacs,nofootinbib]{revtex4-1}
\pdfoutput=1
\usepackage{graphicx}
\usepackage{longtable}
\usepackage{float}
\usepackage{dcolumn}
\usepackage{bm}
\usepackage{appendix}
\usepackage{multirow}
\usepackage{color}
\usepackage[utf8]{inputenc}
\usepackage{footmisc}
\usepackage{soul}
\usepackage{txfonts}
\usepackage{xcolor}
\usepackage{graphicx}
\usepackage{bm}
\usepackage{ulem}
\usepackage{multirow, array, float, tabularx}

\usepackage{amsmath}

\usepackage[colorlinks=true,linkcolor=blue,citecolor=blue]{hyperref}
\usepackage{feynmp-auto}
\usepackage{fancyhdr}
\usepackage{lineno}

\usepackage{wasysym}

\begin{document}
\title{ Exploring solutions to the muon g-2 anomaly in a 3-3-1 model under flavor constraints}



\author{A. Doff$^a$}
\author{Jo\~ao Paulo Pinheiro$^b$} 
\author{C. A. de S. Pires$^{c}$}
\affiliation{$^{a}$ Universidade Tecnologica Federal do Parana - UTFPR - DAFIS, R. Doutor Washington Subtil Chueire, 330 - Jardim Carvalho,  84017-220, Ponta Grossa, PR, Brazil} 
\affiliation{$^{b}$Departament de F\'isica Qu\`antica i Astrof\'isica and Institut de Ci\`encies del Cosmos, Universitat de Barcelona,
  Diagonal 647, E-08028 Barcelona, Spain}
  \affiliation{$^{c}$Departamento de F\'isica, Universidade Federal da Para\'iba, Caixa Postal 5008, 58051-970, Jo\~ao Pessoa, PB, Brazil}

\date{\today}

\begin{abstract} 
The magnetic moment of the muon can receive significant two-loop contributions from a light pseudoscalar. Notably, the spectrum of scalars of 3-3-1 models include one pseudoscalar. However this scalar spectrum inevitably gives rise to flavor-changing neutral current (FCNC) processes. In this study, we examine, within the 3-3-1 model with right-handed neutrinos, whether such spectrum of scalars  can account for the anomalous magnetic moment of the muon, considering the constraints imposed by $B$-meson decays, meson mixing, and invisible Higgs decays.  Our principal finding reveals that a pseudoscalar with  mass around 66 GeV and  $\tan \beta =58$  can account for the $g-2$ anomaly without conflicting with flavor physics.
\end{abstract}

\maketitle
\flushbottom

\section{Introduction}

The anomalous magnetic moment of the muon, $a_\mu \equiv (g_\mu -2)/2$, stands amongst the best probes of the Standard Model (SM) of particle physics and its possible extensions. Recently, the Muon $g$$-$$2$ Experiment at Fermilab announced a new experimental (exp) measurement of $a_\mu$~\cite{Muong-2:2023cdq,Muong-2:2024hpx} which, when combined with the previous and consistent results from the same experiment~\cite{Muong-2:2021ojo,Muong-2:2021xzz,Muong-2:2021ovs,Muong-2:2021vma} (and from the earlier Muon $g$$-$$2$ Experiment at Brookhaven~\cite{Muong-2:2006rrc,Muong-2:2002wip,Muong-2:2004fok}), results in a new world average of $a^{\rm exp}_\mu = 116592059(22)\times 10^{-11}$ with an unprecedented precision of 190 parts-per-billion (ppb). The standard model prediction, when evaluated using data driven dispersive approach, provides $a_\mu^{\rm SM}=116591810(43)\times 10^{-11}$ \cite{Aoyama:2020ynm}, which implies in  $\Delta a_\mu \equiv a^{\rm exp}_\mu-a^{\rm SM}_\mu=(24.9 \pm 4.8) \times 10^{-10}$, yielding a discrepancy of $5.1 \sigma$\footnote{A new $g-2$ experiment with a completely different approach is currently under construction at J-PARC\cite{Abe:2019thb}}.  The explanation of such  discrepancy suggests the necessity of new physics beyond the Standard Model (SM). The task of theoretical particle physics is to find scenarios of new physics that provide an explanation to such discrepancy. Generally, any new physics contributing to $a_\mu$ likely involves hypothetical new particles, including vector or scalar bosons, fermions, or pseudoscalars. Some of these particles may contribute positively to $a_\mu$, while others may contribute negatively\footnote{However, it is important to stress that high-precision lattice QCD calculations  result in independent, but consistent, values for $a_\mu$ that are in agreement with $a_{\mu}^{\rm exp}$ and support a no-new-physics  in the muon $g-2$ scenario\cite{Aoyama:2020ynm}}.

The challenge in resolving the $a_\mu$ anomaly arises from the difficulty in balancing the positive and negative contributions to align with the observed value of $a_\mu$. Among the various new contributions, the one from pseudoscalars is singular because, at the one-loop level, it contributes negatively, whereas at the two-loop level, it contributes positively\footnote{The contribution of  charged scalars is negative for one-loop and positive for two-loops too, however (in the discussed scenario) as we will present in the text, $m_{h_1^+} \geq 483$ GeV, leading to a subdominant contribution (in comparison to $A$)}. Thus, explaining $a_\mu$ solely with pseudoscalars requires a pseudoscalar with a mass in the tens of GeVs\cite{Broggio:2014mna,Han:2015yys}. Consequently, scenarios featuring pseudoscalars that can accommodate $a_\mu$ are highly motivated for further investigation\cite{Wang:2014sda,Abe:2015oca,Crivellin:2015hha,Chun:2016hzs,Han:2015yys,Ilisie:2015tra,Cherchiglia:2018acu}.

Models based on the $SU(3)_C \times SU(3)_L \times U(1)_N$ (3-3-1) symmetry present an intriguing avenue for exploring  physics beyond the SM. These models possess a large  spectrum of particles contributing to $a_\mu$. However the extensive investigation done in Ref. \cite{deJesus:2020ngn}  showed that  no one version of 3-3-1 models solves the current muon $g-2$ anomaly \footnote{The simplest extension of the 3-3-1 model with right-handed neutrinos  that explains $\Delta a_\mu$ at one loop level involves the introduction of a singlet of scalar leptoquark\cite{Doff:2024cap}. For other possibilities, see Refs. \cite{Hong:2024yhk,Pinheiro:2021mps,deJesus:2020ngn,Kelso:2014qka,Ky:2000ku,CarcamoHernandez:2020pxw}}. This finding was based on an energy regime where all the 3-3-1 models belongs to the TeV scale and only single-loop contributions were considered. The motivation for the current study is to explore the muon $g-2$ anomaly within a version of the 3-3-1 
 models called the 3-3-1 model with right-handed neutrinos (331RHN)\cite{Foot:1994ym,Montero:1992jk}. The study will be performed  in a regime of energy where some neutral components of the spectrum of scalars of the model, as its intrinsic pseudoscalar, acquire masses at electroweak scale or lower\cite{Pinheiro:2022bcs}  and take into account both one- and -two loop contributions under the constraints of flavor physics.


This work is organized in the follow way: in Sec. II we present  the 331RHN and its variants. In the Sec. III we calculate the one- and two-loops contributions to $\Delta a_\mu$. In Sec. IV we obtain the theoretical and experimental constraints on the masses of the scalars that contribute to the $g-2$. In Sec. V we present our numerical analysis and in Sec. VI we present a short discussion followed by the main conclusions.

\section{The 331RHN  and its variants}
Our study here is restricted to the 331RHN. Each version of 3-3-1 models dismembers into three variants \cite{Oliveira:2022vjo,Oliveira:2022dav}. In order to understand this, we make a short review how this occurs inside the 331RHN.

In the 331RHN the right-handed neutrinos compose the third components of the triplet of leptons,$
f_{l_L}= (
\nu_{l_L}\,\,
e_{l_L}\,\,
\nu^{c}_{l_R})^T$ 
with $l=e\,,\,\mu\,\,\tau$ representing the three SM generations of leptons.

The gauge sector of the model involves the standard gauge bosons plus one  $Z^{\prime}$, two new charged gauge bosons $W^{\prime \pm}$ and two non-hermitian neutral gauge bosons $U^0$and $U^{0 \dagger}$\cite{Long:1996rfd}. All these new gauge bosons must develop mass at TeV scale and due to this their contributions to the $a_\mu$  are negligible as showed in Ref.\cite{deJesus:2020ngn}.

The scalar sector  is composed by the three triplets of scalars $\eta=(\eta^0 \,\,
\eta^- \,\,
\eta^{\prime 0})^T$, $\rho=(\rho^+ \,\,
\rho^0 \,\,
\rho^{\prime +})^T$ and $\chi=(\chi^0 \,\,
\chi^{-} \,\,
\chi^{\prime 0})^T$ with $\eta$ and $\chi$ transforming as $(1\,,\,3\,,\,-1/3)$
and $\rho$ as $(1\,,\,3\,,\,2/3)$\cite{Long:1997vbr}.  After spontaneous breaking of the symmetries, such content of scalar generates masses for all massive particles of the model including fermions and gauge bosons\footnote{For the development of the other sectors of the model, see Refs.\cite{Long:1995ctv,Long:1997hwz,Cao:2016uur,Oliveira:2022vjo}}. In order to avoid spontaneous breaking of the lepton number, we assume that only  $\eta^0\,\,, \rho^0$, and $\chi^{\prime 0}$ develop VEV, namely $v_\eta\,,\,v_\rho\,,\,v_{\chi^{\prime}}$ respectively 
 \footnote{For the case where the other neutral scalars develop VEVs, see Ref. \cite{deSPires:2003wwk}}. The potential, the minimum conditions and the spectrum of scalars  of the model are found in Ref. \cite{Pinheiro:2022bcs}.  We just remember here that  $v^2_\eta + v^2_\rho= v^2$ where $v$  is the standard vev whose value is $246$ GeV. The vev, $v_{\chi^{\prime}}$, characterizes the energy scale of the breaking of the $SU(3)_L \times U(1)_N$ symmetry.  Current LHC bounds impose  $v_{\chi^{\prime}} \geq 10$TeV\cite{Coutinho:2013lta,Alves:2022hcp}. Moreover, we remember that the potential of the model involves a trilinear term modulated by the energy parameter $f$. This parameter is very important because it regulates the range of values   the masses of the spectrum of scalar may achieve. It was showed in Ref. \cite{Pinheiro:2022bcs} that when  $f< v_\eta\,,\, v_\rho$ the range of values of the masses of the scalars may vary from tens of GeVs up to TeV scale . For $f\geq v_\eta\,,\, v_\rho$ all the spectrum of new scalars, except the Higgs, develops mass at $v_{\chi^{\prime}}$ scale.  Our interest here is  exclusively in the first  case in which $f < v_\eta, v_\rho,v_{\chi^{\prime}}$. In general the spectrum of scalars of the model is composed by three CP-even scalars  $h_1$,$h_2$ and $ H$, two  charged scalars $ h^+_1$ and $h^+_2$, the pseudo scalar $A$  and one neutral bilepton $\eta^{ \prime 0}$. The scalar $h_1$ will play the role of the standard-like Higgs. The pseudo scalar $A$ has mass proportional to the trilinear coupling $f$, meaning it may be  light if $f$ is  small.  What we do here is to investigate the contributions of this set of scalars to the muon ($g-2$) at one and two loops in a general way taking into account the contributions of the gauge bosons of the 331RHN inside the variant of the model that enhance the couplings of the charged leptons with the pseudo scalar $A$ by means of $\tan \beta$.  In other words, we complement the job done in Ref. \cite{deJesus:2020ngn}.

We remember here that in 331  models   anomaly cancellations demand that one family of quarks transforms differently from the other two ones. This feature generates the variants of the models\cite{Oliveira:2022vjo}. In what follow we consider the three variants and separate the main aspects of them that matter to the calculation of their contributions to the $a_\mu$  at one and two loops.

\subsection{Variant I}
In this variant the first two families of quarks  transform as anti-triplet while the third one  transforms as triplet by $\text{SU}(3)_\text{L}$,
\begin{eqnarray}
&&Q_{i_L} = \left (
\begin{array}{c}
d_{i} \\
-u_{i} \\
d^{\prime}_{i}
\end{array}
\right )_L\sim(3\,,\,\bar{3}\,,\,0)\,,u_{iR}\,\sim(3,1,2/3),\,\,\,\nonumber \\
&&\,\,d_{iR}\,\sim(3,1,-1/3)\,,\,\,\,\, d^{\prime}_{iR}\,\sim(3,1,-1/3),\nonumber \\
&&Q_{3L} = \left (
\begin{array}{c}
u_{3} \\
d_{3} \\
u^{\prime}_{3}
\end{array}
\right )_L\sim(3\,,\,3\,,\,1/3),u_{3R}\,\sim(3,1,2/3),\nonumber \\
&&\,\,d_{3R}\,\sim(3,1,-1/3)\,,\,u^{\prime}_{3R}\,\sim(3,1,2/3),
\label{quarks} 
\end{eqnarray}
where  the index $i=1,2$ is restricted to only two generations. The negative signal in the anti-triplet $Q_{i_L}$ is just to standardise the signals of the charged current interactions with the gauge bosons.  The primed quarks are new heavy quarks with the usual $(+\frac{2}{3}, -\frac{1}{3})$ electric charges. 

Here the simplest  Yukawa interactions that generate the correct mass for all standard quarks are composed by the terms\footnote{For the most general Yukawa interactions involving terms that violate lepton number, see: \cite{Doff:2006rt}.},
\begin{equation}\label{yukawa}
-{\cal L}_Y \supset g^1_{ia} \bar Q_{i_L} \eta^* d_{a_R} + h^1_{3a} \bar Q_{3_L} \eta u_{a_R} + g^1_{3a} \bar Q_{3_L} \rho d_{a_R} + h^1_{ia} \bar Q_{i_L} \rho^* u_{a_R} + \mbox{H.c.}\,,
\end{equation}
where $a=1,2,3$ and the parameters $g^1_{ab}$ and $h^1_{ab}$ are Yukawa couplings that, for sake of simplification,  we consider reals. Observe that in this variant the dominant term that determines the mass of the quark top involves  $v_\eta$. Than, in this case, it is natural to assume $v_\eta > v_\rho$. With all this in hand, after developing the Yukawa and the scalar sectors as in Refs.\cite{Cherchiglia:2022zfy,Fan:2022dye,Okada:2016whh}, we present the Yukawa interaction terms that give the main contributions  to the $g-2$ of the muon at two loops,
\begin{eqnarray}
 {\cal L}_Y&=& i \left( -\frac{\tan{\beta}}{v} (V^u_L)_{i3} (V^u_L)_{3i} + \frac{\cot{\beta}}{v}(V^u_L)_{33} (V^u_L)_{33} \right)m_t \bar t \gamma_5 t A \nonumber \\
&+&i \left( -\frac{\cot{\beta}}{v} (V^d_L)_{i3} (V^d_L)_{3i} + \frac{\tan{\beta}}{v}(V^d_L)_{33} (V^d_L)_{33} \right)m_b \bar b \gamma_5 b A \nonumber \\
&+&i \frac{tg \beta}{v}m_l \bar l \gamma_5 l A \nonumber \\
\label{YcaseII}
\end{eqnarray}
where the subscripts  $i=1,2$. The charged leptons are represented by $l=e\,,\, \mu\,,\, \tau$. The standard vev is $v=247$GeV and $V^{u,d}_L$ are the  matrices that mix the left handed quarks. Right-handed quarks are assumed in a diagonal basis . It this variant we define $\tan{ \beta} =\frac{v_\eta}{v_\rho}>1$.


\subsection{Variant II}
In this case the first family transforms as triplet and the second and third families transform as anti-triplet, which means
\begin{eqnarray}
&&Q_{1_L} = \left (
\begin{array}{c}
u_{1} \\
d_{1} \\
u^{\prime}_{1}
\end{array}
\right )_L\sim(3\,,\,3\,,\,1/3)\,,u_{1R}\,\sim(3,1,2/3),\,\,\,\nonumber \\
&&\,\,d_{1R}\,\sim(3,1,-1/3)\,,\,\,\,\, u^{\prime}_{iR}\,\sim(3,1,2/3),\nonumber \\
&&Q_{iL} = \left (
\begin{array}{c}
d_{i} \\
-u_{i} \\
d^{\prime}_{i}
\end{array}
\right )_L\sim(3\,,\,\bar 3\,,\,0),u_{3R}\,\sim(3,1,2/3),\nonumber \\
&&\,\,d_{3R}\,\sim(3,1,-1/3)\,,\,d^{\prime}_{3R}\,\sim(3,1,-1/3),
\label{quarks_1} 
\end{eqnarray}
where  the index $i=2,3$ is restricted to only two generations.

The minimal set of Yukawa interactions that leads to  the correct quark masses involves the terms,
\begin{eqnarray}
&-&{\cal L}_Y \supset  g^2_{1a}\bar Q_{1_L}\rho d_{a_R}+ g^2_{ia}\bar Q_{i_L}\eta^* d_{a_R} \nonumber \\
&&+h^2_{1a} \bar Q_{1_L}\eta u_{a_R} +h^2_{ia}\bar Q_{i_L}\rho^* u_{a_R} + \mbox{H.c.}\,.
\label{yukawa1}
\end{eqnarray}
 Observe that in this case the main contribution to the mass of the quark top is determined by the vev $v_\rho$. Then in this case we assume  $v_\rho > v_\eta$.
 
Following all the previous procedure done in Variant I, we obtain the following Yukawa interactions involving $A$ and the standard charged fermions that matter to the calculation of the $g-2$ of the muon at two loops,
\begin{eqnarray}
 {\cal L}_Y&=& i \left( \frac{\tan \beta}{v} (V^u_L)_{13} (V^u_L)_{3} - \frac{\cot \beta}{v}(V^u_L)_{i3} (V^u_L)_{3i} \right)m_t \bar t \gamma_5 t A \nonumber \\
&+&i \left( \frac{\cot \beta}{v} (V^d_L)_{13} (V^d_L)_{31} - \frac{\tan \beta}{v}(V^d_L)_{i3} (V^d_L)_{3i} \right)m_b \bar b \gamma_5 b A \nonumber \\
&+&\frac{i}{v \tan \beta}m_l \bar l \gamma_5 l A \nonumber \\
\label{YcaseII_2}
\end{eqnarray}
where the subscripts  $i=2,3$. Here we  define $\tan \beta =\frac{v_\rho}{v_\eta}>1$.


\subsection{Variant III}

In this case the second family transforms as triplet while the first and third ones transform as anti-triplet
\begin{eqnarray}
&&Q_{2_L} = \left (
\begin{array}{c}
u_{2} \\
d_{2} \\
u^{\prime}_{2}
\end{array}
\right )_L\sim(3\,,\,3\,,\,1/3)\,,u_{2R}\,\sim(3,1,2/3),\,\,\,\nonumber \\
&&\,\,d_{2R}\,\sim(3,1,-1/3)\,,\,\,\,\, u^{\prime}_{2R}\,\sim(3,1,2/3),\nonumber \\
&&Q_{iL} = \left (
\begin{array}{c}
d_{i} \\
-u_{i} \\
d^{\prime}_{i}
\end{array}
\right )_L\sim(3\,,\,\bar 3\,,\,0),u_{iR}\,\sim(3,1,2/3),\nonumber \\
&&\,\,d_{iR}\,\sim(3,1,-1/3)\,,\,d^{\prime}_{iR}\,\sim(3,1,-1/3),
\label{quarks_2} 
\end{eqnarray}
where $i=1,3$.

Here the Yukawa interactions involve the terms
\begin{eqnarray}
&-&{\cal L}_Y \supset g^3_{2a}\bar Q_{2_L}\rho d_{a_R}+ g^3_{ia}\bar Q_{i_L}\eta^* d_{a_R} \nonumber \\
&&+h^3_{2a} \bar Q_{2_L}\eta u_{a_R} +h^3_{ia}\bar Q_{i_L}\rho^* u_{a_R} + \mbox{H.c.},
\label{yukawa1_2}
\end{eqnarray}
where $i=1,3$.

Following the procedure of Variant I the interactions of the pseudoscalar $A$ with the standard quarks and leptons that matter to the calculation for the $g-2$ of the muon at two loops are given by,
\begin{eqnarray}
 {\cal L}_Y&=& i \left( \frac{\tan{\beta}}{v} (V^u_L)_{23} (V^u_L)_{32} - \frac{\cot\beta}{v}(V^u_L)_{i3} (V^u_L)_{3i} \right)m_t \bar t \gamma_5 t A \nonumber \\
&+&i \left( \frac{\cot{\beta}}{v} (V^d_L)_{23} (V^d_L)_{32} - \frac{\tan{\beta}}{v}(V^d_L)_{i3} (V^d_L)_{3i} \right)m_b \bar b \gamma_5 b A \nonumber \\
&+&i \frac{m_l}{v \tan{\beta}}\bar l \gamma_5 l A \nonumber \\
\label{YcaseII_3}
\end{eqnarray}
where   $i=1,3$, $v=247$GeV and $V^{u,d}_L$ are the quark mixing matrices. Here we define  $\tan \beta =\frac{v_\rho}{v_\eta}>1$.


 All this tell us that each variant of the model will lead to different physical results in what concern flavor physics. We stress that the  differences among the three variants that matter to muon $g-2$ are the Yukawa interactions of $A$ with the charged leptons $l$. In the case of variant II and III these interactions are suppressed by $\tan \beta$ while in the variant I it is enhanced by $\tan \beta$.  So we deduce that only the variant I has potential to explain $\Delta a_\mu$ at two loops. It is for this reason that from now on we focus exclusively in the variant I of the 331RHN. Observe that we presented the variants involving only  the fermions that contribute to  $a_\mu$. In the APPENDIX we present the general Yukawa interactions involving  all the set of scalars for the case of variant I that is the interesting one. 
  
\section{\texorpdfstring{Two loops contributions to $\Delta a_\mu$ in the 331RHN}{2loopg-2}}

\begin{figure}
    \centering
\includegraphics[scale=0.5]{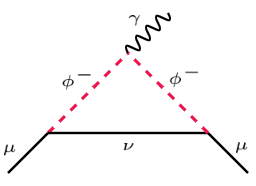}
\includegraphics[scale=0.5]{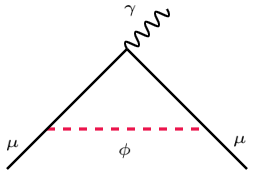}
\includegraphics[scale=0.5]{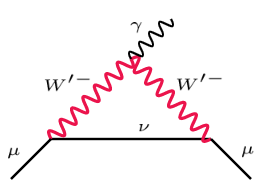}
\includegraphics[scale=0.5]{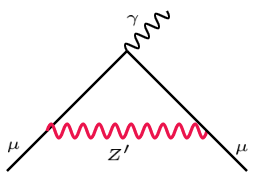}
    \caption{One-loop contributions to the $g-2$ of the muon in the 331RHN where $\phi\,\,\mbox{and}\,\, \phi^-$  represent the neutral and charged scalar contributions}
    \label{fig:pseudoscalarV1_2}
\end{figure}

Before discussing the two loop case, we first review the main aspects of one loop contributions to $a_\mu$ inside the 331RHN. One loop contributions to the muon anomalous magnetic moment in the 331RHN was extensively investigated in \cite{deJesus:2020ngn}. There it was assumed that  $\tan \beta=1$ and that the mixing matrices $V^{u,d}_L$ follow the hierarchy of the CKM one with the main contributions being proportional to $(V^{u,d}_L)_{33}$ which was taken to be $\approx 1$. In this case it does not make sense to talk about variants because all the three variants will give the same contributions. In this case the  main contributions to $\Delta a_\mu$ at one loop are due to the new gauge bosons $W^{\prime \pm}$ and $Z^{\prime}$, the single charged scalar $h^+_1$, the neutral scalar $h_2$, and the pseudo scalar $A$. The set of interactions and their respective contributions to $\Delta a_\mu$  are given in Refs. \cite{Lindner:2016bgg,deJesus:2020ngn}. The Feymnan diagrams  due to  these  contributions  are showed in FIG. \ref{fig:pseudoscalarV1_2}.

 As it was extensively discussed in Ref.\cite{deJesus:2020ngn},  although the  331RHN yields five new contributions at one loop to g-2 with   $W^{\prime}$ and $Z^{\prime}$ giving the dominant positive contributions, unfortunately these five contributions are not able to accommodate the current value of  $a_\mu$ because the negative contributions given by the scalars of the model surpass the positive ones. Thus if we want the 331RHN  explains $\Delta a_\mu$ we need to find positive contributions that   surpass the negative ones given at one loop.

It is very well known that $A$ is responsible for the dominant contribution in two-loops to  $a_\mu$\cite{Cherchiglia:2016eui},\cite{Chang:2000ii}, \cite{Chun:2019oix} and \cite{Ferreira:2021gke}. The 331RHN has a pseudoscalar, $A$, in its spectrum of scalars. We then ask for what range of values for $m_A$ the contributions of the set of scalars ($h_1\,,\,h_2\,,\, h^+_1\,,\, A$) in two loops  would accommodate the current value of  $\Delta a_\mu$. In the two loops, contributions involving $h^+_1$, $W^{\prime +}$ and $Z^{\prime}$ are very suppressed. We neglect them and consider only contributions involving  $A$ and $h_2$. We stress that the contribution of $h_2$, although small, it is  relevant in obtaining the current value of $\Delta a_\mu$ according to dispersion methods.
\begin{figure}
    \centering
\includegraphics[scale=3.1]{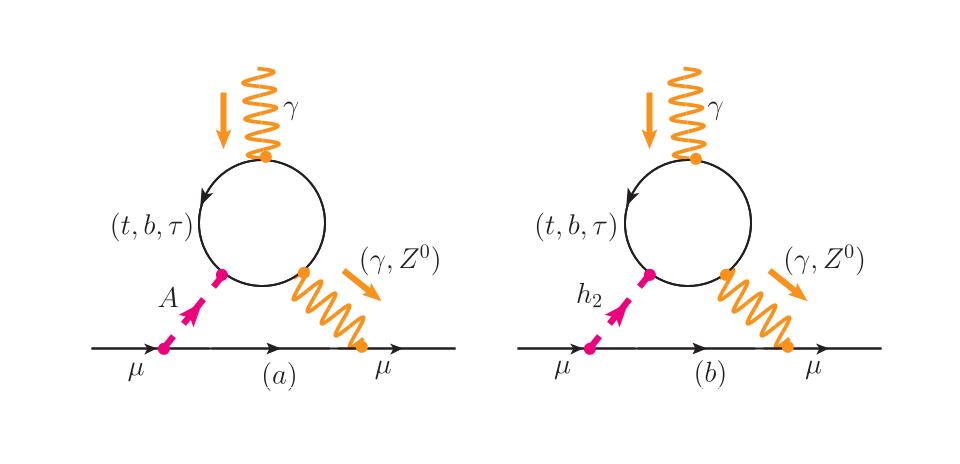}
    \caption{ Two loops Barr-Zee type  contributions to the $g-2$ of the muon in the 331RHN. In this figure 
    $h_2$ and  $A$  represent the neutral scalar and 
    pseudoscalar  contributions  to $g-2$  of the muon. 
    }
    \label{twoloop}
\end{figure}

\begin{figure}
    \centering
\includegraphics[scale=0.5]{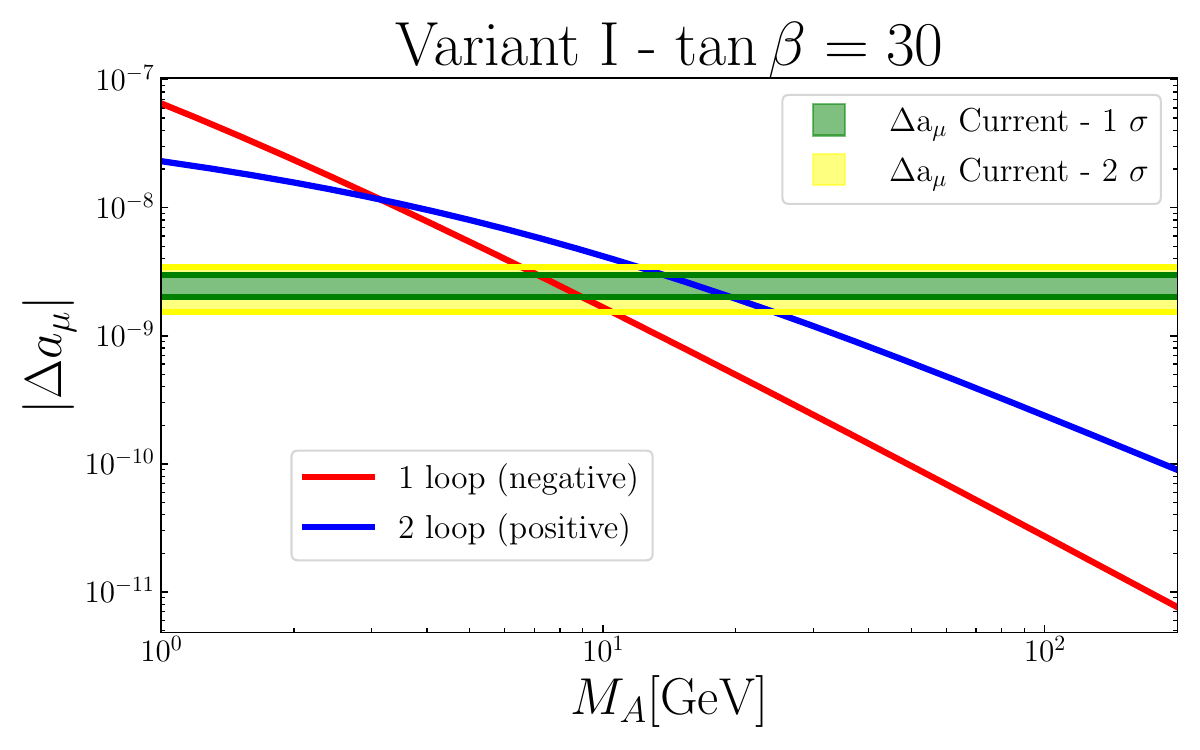}
    \caption{One and two-loops pseudoscalar contributions for the anomalous g-2 of the muon. The green band represents the actual bounds for the anomaly, the red curve represents the module of the one-loop contribution and the blue curve represents the two-loop contribution of the pseudoscalar A.  }
    \label{fig:one-two-loop}
\end{figure}

\begin{figure}
    \centering
\includegraphics[scale=0.5]{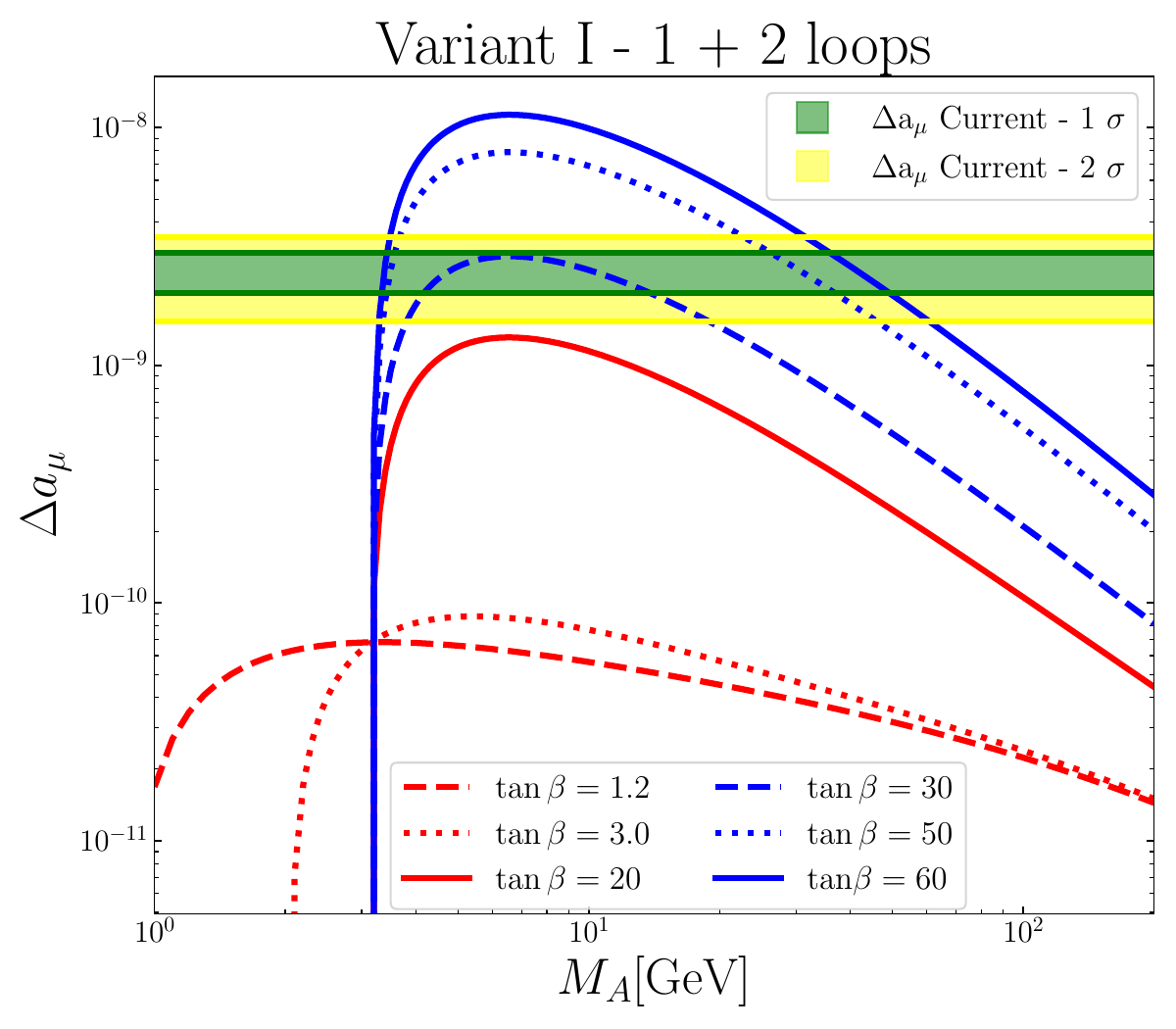}
    \caption{Here, the curves represents the sum of one and two-loop contributions for the anomalous magnetic moment of the muon for a fixed $\tan \beta$. It is clear that, for $\tan \beta >27$, it is always possible to find a point in $[M_A,\tan \beta]$ plane that explains the $\Delta a_\mu$. For more details, see the Appendix.}
    \label{fig:assimptoticlimit}
\end{figure}

Following \cite{Chang:2000ii}, \cite{Chun:2019oix}, \cite{Ferreira:2021gke} and \cite{Cherchiglia:2016eui}, we will discus the contributions to  $ a_\mu$ from one and two loops. The expressions for the contributions of one loops are found  in Refs. \cite{Lindner:2016bgg} and \cite{deJesus:2020ngn}. For the case of the two loops with $A$ and $h_2$ we have two dominant contributions:  one involving  $\gamma$ and other involving $Z$-boson. The expressions for these contributions are found in Refs. \cite{Chang:2000ii}, \cite{Chun:2019oix} and \cite{Ferreira:2021gke} .



In what follow we evaluate such contributions for the  variant I of the model as a function of $M_A$ and $\tan \beta$\footnote{Here we refer to 331RHN because in one loop we consider the main contribution due to $W^{\prime +}$.}. In FIG. (\ref{twoloop}) we show the Feymnan diagrams of the two loops contributions and  in FIG. (\ref{fig:one-two-loop}) we present our results for the cases of  one and two loops. In that figure  the  red curve represents the module of the one loop contributions while blue curve represent the two loops contributions due to the pseudo scalar $A$, only, that is definitively positive.  The green band represents the current bound for the anomaly. The intersection point of the blue and red curves represent the point the two loops contributions cancel the net one loop negative contributions. It occurs at $M_A \approx 3$ GeVs. From that point ahead the two loops start giving  a net positive contribution to $\Delta a_\mu$.  The intersection point of the blue curve with the green band occurs for $M_A$ in the range $15-25$ GeV. Thus the result presented in FIG. (\ref{fig:one-two-loop}) tell us that for $A$ relatively light its two loops contribution surpasses the negative one loop contributions of the spectrum of 331  model that contributes to muon $g-2$ and provides the current $\Delta a_\mu$. The value of $M_A$ in this plot may increase a little  as function of $\tan \beta$. We show this in FIG. (\ref{fig:assimptoticlimit}) where we present the behaviour of $\Delta a_\mu$ in function  of $M_A$ for various values of $\tan \beta$ considering one and two loops contributions. We have that for $M_A> 3$ GeV and $\tan \beta> 27$ the 331RHN always offers a point  in the plane $(M_A\,,\, \tan \beta)$  that accommodates the current value of $\Delta a_\mu$.

However to complete our understanding it is recommended to certify what range of values for $M_A$ and $\tan \beta$ accommodates $\Delta a_\mu$ and respect  constraints from flavor physics. More precisely, we must certify the contributions of $A$ to $K^0 -\bar K^0$ and B mesons decay, in addition to the decay $h_1 \to AA$. We discuss all these  constraints right below.


\section{Theoretical and experimental limits}

\subsection{\texorpdfstring{$K^0-\bar K^0$}{K0-K0bar}}
\par In  the 331RHN  the set of neutral particles $h_1\,,\,h_2\,,\,H\,,\,A$ and $Z^{\prime}$ contribute to FCNC processes\footnote{For important works treating flavor physics in 331  models, see Refs. \cite{Promberger:2007py,Benavides:2009cn,Buras:2012dp, Buras:2013dea, Buras:2023ldz,NguyenTuan:2020xls}}. In the scenario where $A$ solves the muon $g-2$ anomaly, where it is the lightest of this set of neutral particles, we must worry with its contribution to FCNC processes as meson transitions. Then it makes mandatory to certify if a pseudoscalar with mass of tens of GeVs is in agreement, for example, with the current value of $\Delta m_K$ associated to the $K^0 - \bar K^0$ transition.  In this regard,  the   effective Lagrangian  that leads to the $K^0 - \bar K^0$ transition can be write as  \cite{Oliveira:2022vjo}

\begin{equation}
{\cal L^K_{\rm eff}} = -\frac{1}{m^2_A}\left[\bar{d}\left(C^{R}_KP_R - C^{L}_KP_L\right)s \right]^2 ,  
\label{LDK}
\end{equation}

\noindent where in the equation above $P_{L,R}$ are  the left-handed and right-handed projections. In our case the coefficients  $C^{R,L}_K$ are obtained from the Yukawa interactions given in  Eq.(\ref{YcaseIA}) of  the APPENDIX. With those interactions we can write  

\begin{eqnarray}
&& C^{R}_K = \left(\tan \beta (V^d_L)_{32}(V^d_L)_{13} +  \cot \beta(V^d_L)_{i2}(V^d_L)_{1i} \right)\frac{m_s}{v}\nonumber\\
&&C^{L}_K = \left(\tan \beta (V^d_L)^*_{13}(V^d_L)^*_{32} + \cot \beta(V^d_L)^*_{1i}(V^d_L)^*_{i2} \right)\frac{m_d}{v},
\label{eqCK1}
\end{eqnarray}
\noindent where $i=1,2$.

\par  With the  Lagrangian in Eq.(\ref{LDK}), the coefficients in Eq. (\ref{eqCK1}) and following the procedure described in 
\cite{Oliveira:2022vjo},  we obtain the following expression to the mixing parameter  $\Delta m_K$ as function of the coefficients  $C^{R,L}_K$
\begin{equation}
  \Delta m_K =  \frac{2m_Kf^2_k}{m^2_A}\left(\frac{5}{24}Re\left ((C^{L}_K)^2+(C^{R}_K )^2\right)F^2(k)
   + 2Re\left(C^{L}_KC^{R}_K\right)G^2(k)\right),
   \label{eqDK}
\end{equation}
\noindent where 
\begin{eqnarray}
&&F^2(k) = \left(\frac{m_K}{m_s + m_d}\right)^2 ,\nonumber \\
&&G^2(k) = \left(\frac{1}{24}+\frac{1}{4}F^2(k)\right)\nonumber.  
\label{eFG}
\end{eqnarray}
\noindent In the Eq.(\ref{eqDK}) $f_K$ is the kaon($K^0$) decay constant and $m_k$ is the kaon  mass. The  input parameters that will be considered in these equations are listed  below
\begin{eqnarray}
&&  m_s = 93.4\,\mathrm{MeV}\,\,,\,\, m_d = 4.67\,\mathrm{MeV}   \nonumber \\
&&  m_K = 497.6\,\mathrm{MeV} \,\,,\,\, f_K =  156\,\mathrm{MeV}  \nonumber.
\end{eqnarray}
\par  The coefficients $C^{R,L}_K$ depend on the mixing matrices $V^d_L$ and $V^u_L$ whose entries are unknown free parameters that obey the constraint,
\begin{equation}
V^u_LV^{d\dagger}_L = V_{CKM},
\label{ckm}
\end{equation}
which is not sufficient to fix all the entries of $V^{u,d}_L$. The pattern of $V^{u,d}_L$ involves a high level of arbitrariness. A  common procedure in works treating meson transitions is to assume aleatory parametrization to the mixing matrix $V^{u,d}_L$, see for example Refs. \cite{Cogollo:2012ek,Queiroz:2016gif,Okada:2016whh, CarcamoHernandez:2022fvl}. However, it was noted in \cite{Oliveira:2022dav} that once the standard-like Higgs contributes inevitably to meson transitions, too, as it is exposed in the Yukawa interactions in Eq. (\ref{Ycaseh1}), and as its contribution depends exclusively of $V^{u,d}_L$ for fixed angles $\beta$ and $\alpha$, it was then realized that the main role of the Higgs  contribution to meson transitions is  to auxiliar  in the choice of the parametrization of the mixing matrices $V^{u,d}_L$ .  


We then use the following parametrization to the  quarks mixing

\begin{equation}
\hspace*{-4cm}
V^d_L\approx
\begin{pmatrix}
0.999988 &  0.004863 & -0.0003028\\
0.004872 & -0.997974 & 0.0634406\\
6.237\times10^{-6}& -0.063441 & -0.997974
\end{pmatrix}
\label{MDKd}
\end{equation}
\begin{equation}
\hspace*{-1cm} 
V^u_L \approx
\begin{pmatrix}
0.975375 - 2.06267\times10^{-8}\,i & -0.220224 + 0.000209784\,i & 0.0127631 + 0.00330004\,i \\
-0.22045 - 0.000137853\,i & -0.975187 - 6.70389\times 10^{-7}\,i & 0.0202181 + 4.17545\times 10^{-8}\,i \\ 
 0.00799242 - 0.00322638\,i& -0.0225194 + 0.00072796\,i & -0.999704 - 0.0000462961\,i
\end{pmatrix}
\label{MDKu}
\end{equation}


 
\noindent In  obtaining this parametrization above, following the procedures in Ref. \cite{Oliveira:2022dav},  we demanded that the contribution of  $h_1$  to the experimental error of $\Delta m_K^{exp}$ is negligigle  and that the contribution of $A$ is smaller than the experimental error. This makes of  $V^{u,d}_L$ in Eqs. (\ref{MDKd}) and Eq. (\ref{MDKu})  a realistic parametrization for the quark mixing. Throughout this paper we use this parametrization in our calculations.

We are now ready to investigate the effects of a light pseudoscalar, $A$, in flavor physics. More precisely, we investigate if the $K^0 -\bar K^0$ transition is compatible with $M_A \sim (60-70)$ GeV's for $\tan \beta \sim 60$ and the pattern of $V^{u,d}_L$ given above. We know that  the Standard Model  contributions to the neutral meson-anti meson transitions   present a good agreement with experiments\cite{Buchalla:1995vs,DiLuzio:2019jyq,DeBruyn:2022zhw}. Furthermore, the  uncertainty due to errors in QCD corrections must be considered\cite{Wang:2019try,Wang:2022lfq}. Therefore, with regard to the contributions from new physics, it is reasonable to consider that these contributions fall inside the experimental error. In the particular case of the  $K^0 - \bar K^0$   meson-anti meson transitions we have,
 $$
 \Delta m_K^\mathrm{exp} = (3.484 \pm 0.006) \times 10^{-12}\,\mathrm{MeV}.
 $$
 \noindent Based on these considerations, the experimental errors $\delta(\Delta m_K)$ of the $K^0 - \bar K^0$ mass difference that we use to constrain new physics is 
\begin{equation}
\delta(\Delta m_K) = \pm 0.006\times 10^{-12}\,\mathrm{MeV} = \pm 0.6\times 10^{-17}\,\mathrm{GeV}.
 \label{KEr}
 \end{equation}
Once we demanded  that the contribution of  $h_1$ to the error is negligible, we then assume that the contribution of the pseudoscalar $A$ to the error  does not surppass $0.995\%$ of the total error\footnote{We stress here that we are aware that the other neutral particles as $h_2$ and $Z^{\prime}$ also contribute to such error. However as they are much heavier than $A$, we think it is reasonable to neglect their contributions to such error\cite{Okada:2016whh,Oliveira:2022dav}.}. Note that with respect to the precision used in the parameterization of the matrices (\ref{MDKd}) and (\ref{MDKu}),  the number of decimal places used in the calculations  was maintained in order to obtain the best result for unitarity. 

\par
\begin{figure}
    \centering
\includegraphics[scale=0.45]{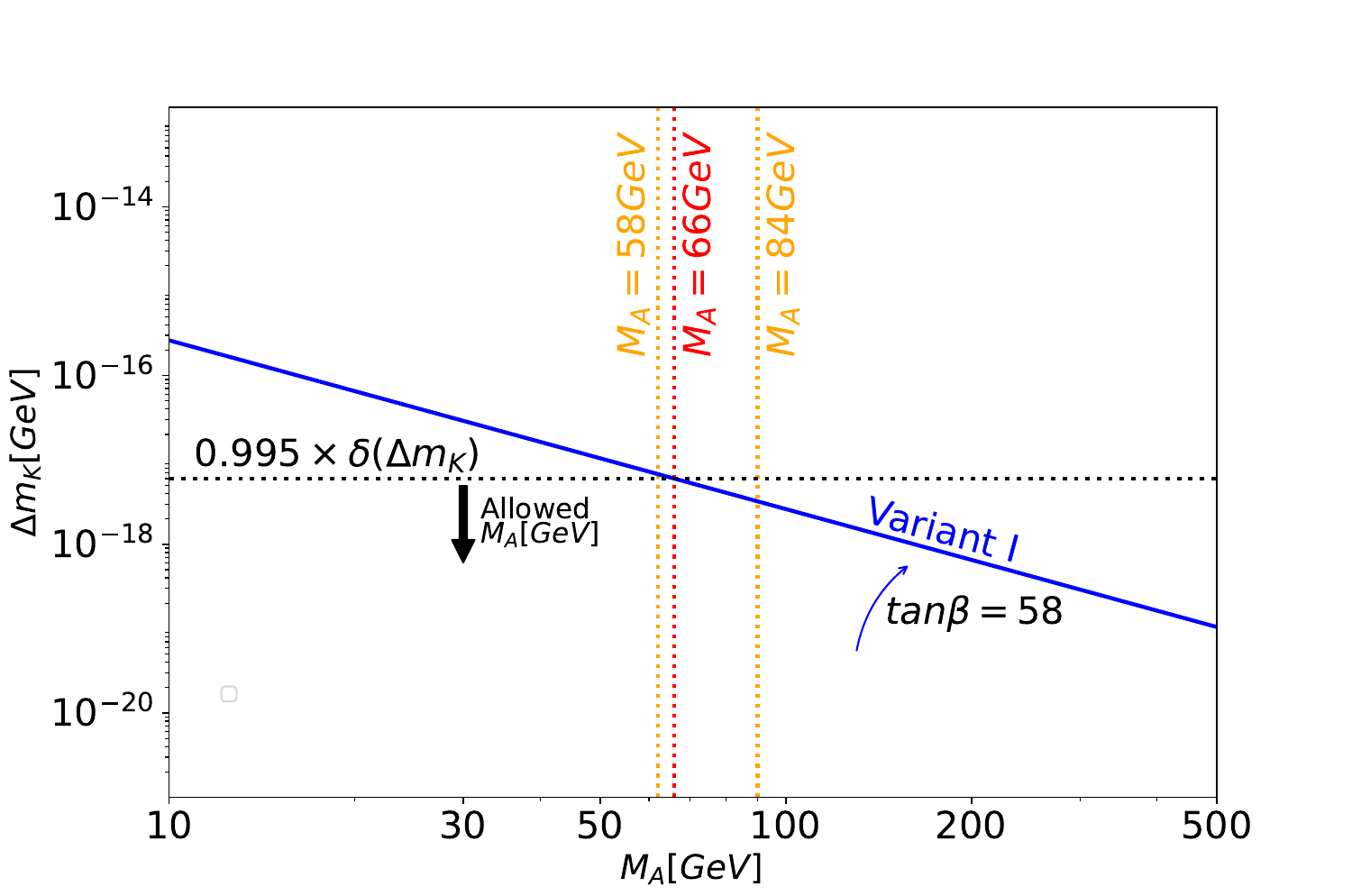}
    \caption{Flavor changing neutral currents constraint on  $m_A$ from  $K^0 - \bar K^0$ transition. We demand  $h_1$ gives negligible contribution to  $\delta(\Delta m_K)$, and assuming that the contribution of the  pseudoscalar $A$ fall inside the error. The contextualization of the curve behavior is described in the text. }
    \label{fig:pseudoscalarV1_1}
\end{figure}

 We present our results in  FIG. (\ref{fig:pseudoscalarV1_1}) where  the blue dashed curve corresponds to the behavior of Eq.(\ref{eqDK}) with $M_A$ under the condition that $\Delta m_K$ stands for $0.995\%$  of the  error and for the pattern of quark mixing described by Eq.(\ref{MDKd}) where $i,j = 1,2,3$. In that figure the black dotted line corresponds to $0.995\%$ of the  error given by Eq.(\ref{KEr}) and the  dashed red vertical line indicates the position in the band where the pseudoscalar mass, $M_A=66$ GeV,  explains the $\Delta a_\mu$ respecting the  constraints by  flavor changing neutral processes.

\subsection{B-meson decays}
In this subsection, we will describe important bounds from flavor observables for the 331RHN
\cite{Deschamps:2009rh,Chowdhury:2017aav,Arnan:2017lxi,Aoki2009,Mahmoudi2009,Jung2010,Buras:2016dxz,Buras:2021rdg}.
\subsubsection{\texorpdfstring{$B_s \to \mu\mu$}{Bsmumu}}

The $B_s \to \mu \mu$ decay suffer from few hadronic uncertainties and
are induced
by FCNC transitions, which make them sensitive probes to the effects
of physics beyond the
SM, especially models with a non-standard Higgs sector. For our
particular case, the process is induced by the gauge bosons $Z$,$W$ and
scalar particles $A,\,h_1,\,h_2$ and $h_1^+$. 
This decay is induced by loop diagrams and any source of FCNC in our theory could in principle contributes to the meson decay. For this analysis, we are 
using the analytical formulas from \cite{Lang:2022mxu} and comparing
with recent bounds from  \cite{HFLAV:2022esi}. One relevant information is that the new vertices that contributes for the diagrams of this decay are proportional to $\cot \beta$, meaning that, for $\tan \beta >1$, one expect that, for the studied case, the contribution from new physics would be marginal.

\subsubsection{\texorpdfstring{$B \to X_s \gamma$}{BXsgamma}}

As constantly discussed in the literature, the inclusive radiative decay $B \to X_s \gamma$ is one of the most important bounds for the mass of a charged scalar \cite{Das:2015qva,Borzumati:1998ie,Ciuchini:1997xe,Arbey:2017gmh,Atkinson:2021eox}
. The meson decay is generated
by the elemental process $b\to s \gamma$ and since the hadron transition occurs in this process, perturbative QCD corrections
are much important and non-perturbative effects must be concerned as well. Here, the Higgs scalar particle gives new contributions to the Wilson coefficients of the effective theory. The process is directly proportional to $\tan \beta$ when $h_1^+$ is light enough, and this decay figures an important way to bound the mass of the scalar $h_1^+$. For sufficiently heavy $h_1^+$, the dependence on $\tan \beta$ vanishes. Then, this bound limits inferiorly the mass of $h_1^+$.

The bounds are enhanced after calculating the diagrams for more loops.
In the literature, there is analytic calculation for NLO \cite{Enomoto:2015wbn}. Recently, a group calculated numerically the NNLO contributions \cite{Misiak:2020vlo} and the constraints on the $m_{h_1}^+$ parameter are extremely strong ($>800$ GeV). However, one important assumption of this paper is that they are not considering FCNC at tree-level.  As discussed in the literature\cite{Branco:2011iw,Mahmoudi2009}, on considering models that allows FCNC, the bounds from the inclusive radiative decay imposed on $m_{h_1}^+$ are relaxed. This is why we are considering in our paper only NLO analytical formulas for the inclusive radiative decay given in \cite{Enomoto:2015wbn} and applying the parameterization developed in the $K-\bar{K}^0$ section. The experimental bounds been used are from \cite{HFLAV:2022esi}.

\subsubsection{\texorpdfstring{$B \to \tau\nu$}{Btaunu}}

Typically there are decay processes like $M^\pm \to \ell
\nu_\ell$ mediated by the charged scalar. They occur at tree-level and
the most relevant bounds comes from the decays $B \to \tau \nu$, $D\to
\mu \nu$, $D_s\to \tau \nu$ and $D \to \nu \mu$. For the considered
case in this article, the strongest contribution is given by the
process $B \to \tau \nu$, and it is very important to obtain a superior
limit for $\tan \beta$ at different $h_1^+$ masses. We performed the analysis applying the
experimental limits from \cite{HFLAV:2022esi} at $95\%$ C.L. and the analytical
formula from \cite{Enomoto:2015wbn}.

\subsection{Higgs decay}

Concerning the Higgs decay, the interaction $
{\cal L}^{hAA} \supset \frac{1}{2} g_{AAh} h_1 A A $
allows the decay mode  $h_1\rightarrow A+A$ when $m_A< m_h/2$. If this is the case, we get
$$\Gamma(h_1 \to AA) = \frac{g_{hAA}^2}{32\pi m_h} \sqrt{1-\frac{4m_A^2}{m_{h_1^2}}}\,\,\,\,\mbox{where}\,\,\, \,\,g_{AAh} = \frac{2v \left(2 \lambda _6 \left(\tan^4\beta+1\right)+\left(\lambda _2+\lambda _3\right) \tan^2\beta\right)}{\left(\tan^2\beta+1\right)^2}.
 $$

This decay enters in the mode invisible of $h_1$. For large values of $\tan \beta$ we have $g_{hAA}\to 4 \lambda_6 v$. Here $\lambda_6$ is a function of the mass of the scalars, 

\begin{eqnarray}
    \lambda_6=\frac{m_{h_1}^2}{v^2}  - \frac{m_{h_2}^2-4 M_A^2 \cot\beta}{v^2},
\end{eqnarray}
and in order to avoid the bounds from the invisible decays of the Higgs, one must artificially reduce the size of $\lambda_6$ to be nearly zero or imposes $m_{h_1}=m_{h_2}$ when $\cot\beta \to 0$. However, without fine tuning the only way to escape from this bound is to impose an inferior limit for $M_A > \frac{m_h}{2}$. Thus we have to explain $\Delta a _\mu$ with $A$ respecting this bound.

\subsection{Instability conditions}
Here we obtain the range of values for the parameters $[M_A \, , \, m_{h^+_1}]$  allowed by the stability conditions of the potential\cite{Haber2015}. For this we use the most general potential of the 331RHN that conserves lepton number which is given in \cite{Pal:1994ba}. In order to avoid spontaneous breaking of the lepton number, it is assumed that only  $\eta^0\,\,, \rho^0$, and $\chi^{\prime 0}$ develop VEV\footnote{For the case where the other neutral scalars develop VEVs, see Ref. \cite{Doff:2006rt}}. 
After diagonalizing the the mass matrices of the scalars of the model, we obtain the tree-level scalars masses\cite{Pinheiro:2022bcs}:
\begin{eqnarray}
 &&M^2 _A=\frac{f v_{\chi^{\prime}}}{4}(\frac{ \tan \beta}{1+\tan \beta^2}\frac{v^2}{v^2_{\chi^{\prime}} }+\tan \beta+\cot \beta)\,\,\,, m_{h_1^{\pm}}^2=\frac{1}{2}(fv_{\chi^{\prime}}+\lambda_9\frac{  \tan \beta}{1+\tan \beta^2} v^2)(\tan \beta+\cot \beta).
 \label{mass}
\end{eqnarray}
 These mass expressions and all the other ones  are found in \cite{Pinheiro:2022bcs}. 
 
For the decoupling limit of the heavy 331 particles  which means to take $v_{\chi^\prime} \geq 10$  TeV, the phenomenological viable set of scalars  are a  mixing of $\rho$ and $\eta$. After the decoupling, the relevant bounds from  below conditions \cite{Barroso:2013awa} are expressed by the set of relations

\begin{eqnarray}
    \lambda_2>0,\,\,\lambda_3>0,\,\, \lambda_6+2\sqrt{\lambda_2 \lambda_3}>0,\,\, \lambda_6+\lambda_9+2\sqrt{\lambda_2\lambda_3}>0,
\end{eqnarray}
while  the unitarity and perturbativity conditions \cite{Ginzburg:2005dt} are given by $|\lambda_i|<4\pi$.

\section{Numerical Analysis}
In this section, we perform a  numerical scan in the parameter space
of the Variant I of the 331RHN. Here, we are assuming the limit $v_\chi\gg v_\eta,v_\rho$. This means
that, in addition to the SM particles, there are only three new
particles, $h_2,\,h_1^\pm$ and $A$. We identify $h_1$ as the standard
Higgs and we are  considering the alignment limit for the CP-even
scalars \cite{Chen:2018shg,Han:2020lta}. For fixed $v_\chi=10$ TeV and $v=246$ GeV,
and adopting the physical basis for the scalar potential in the
331RHN, there are 4 remaining free parameters that we assume varying
in the following range:
$$\{ \tan \beta \in [1,200],m_{h_2} \in [5,1000]\,\mathrm{GeV},M_A
\in [5,1000]\,\mathrm{GeV},m_{h^+_1} \in [100,1000]\,\mathrm{GeV}\}.$$

\begin{figure}
    \centering
    \includegraphics[scale=0.5]{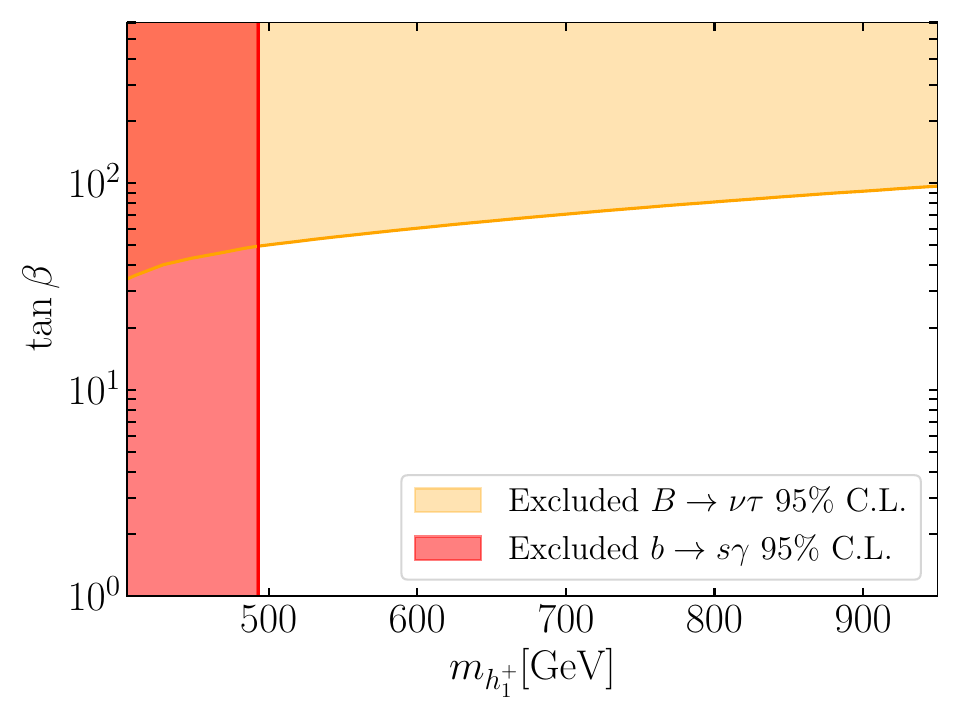}
    \caption{Excluded regions due to $B$-meson decays at 95 $\%$ C.L. in the $[m_{h_1^+},\tan \beta]$ plane. From these bounds, we projected these bounds into the $M_A$, $\tan \beta$ plane without assuming mass degenerescence  between $h_1^+$ and $A$.}
    \label{fig:bounds_hp}
\end{figure}

In FIG. \ref{fig:bounds_hp}, we present 
the constraints from the processes $B\to X_s \gamma$ and $B\to \tau \nu$ in the plane $[ m_{h^+_1}, \, \tan \beta]$. It is clear in that figure  that $m_{h_1^+}$ is bounded from below independently from the value of $\tan \beta$, for $\tan \beta >1$
$$m_{h^+_1}>483\,
\mathrm{GeV},$$ while $B\to \tau \nu$ limits superiorly the value of $\tan \beta$.

 \begin{figure}
    \centering
    \includegraphics[scale=0.5]{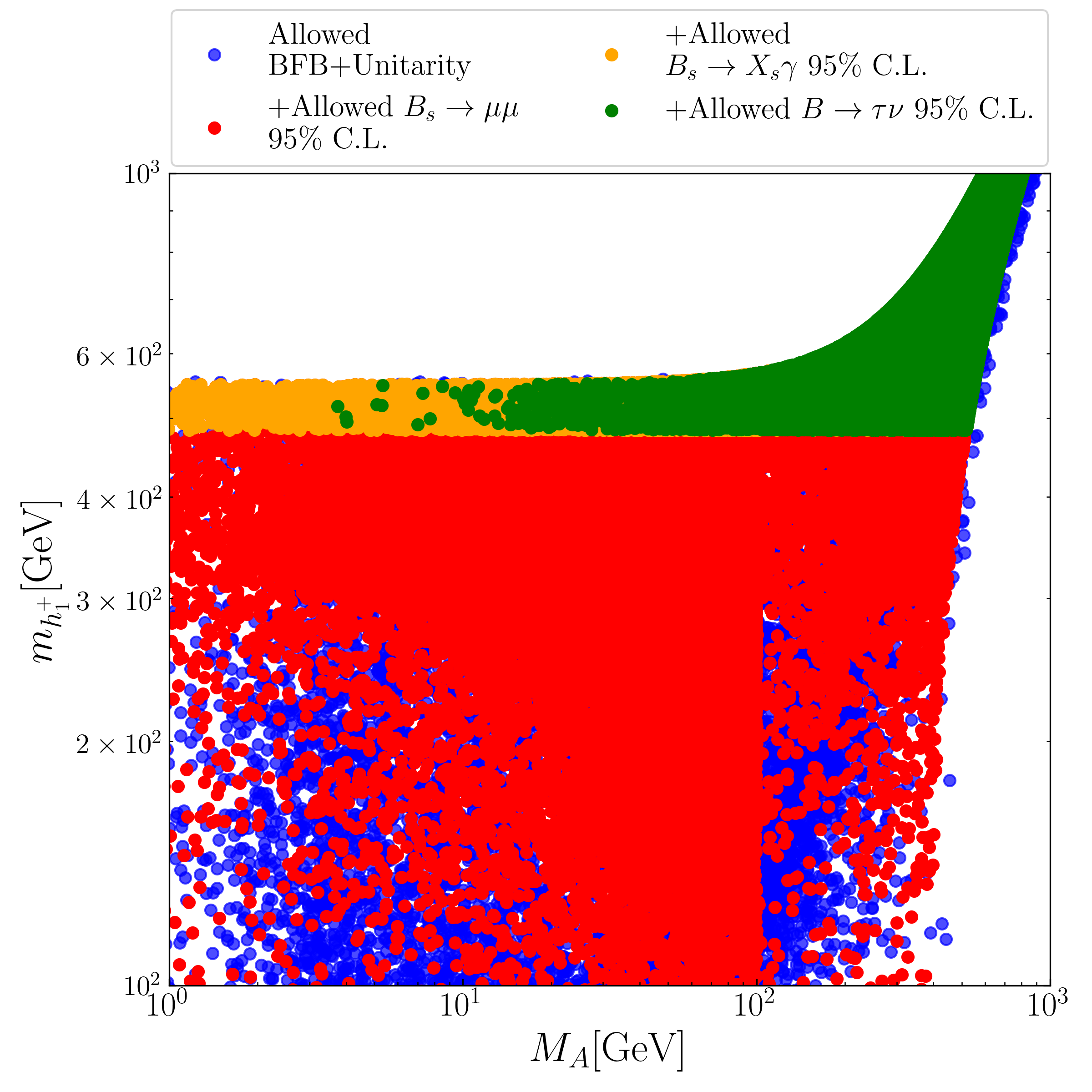}
    \caption{Here, we can see the relation between the masses of the  charged scalar $h^+_1$ and the pseudoscalar $A$. The blue dots represents the possible values for masses in the $[M_A, m_{h^+_1}]$ plane in which the potential bounds are obeyed (bounded from below conditions and imposing that the mass of all scalars are positive), red dots represents the cumulative bounds from the muon decays of $B_s$ meson and potential conditions, the yellow dots represents the simultaneous bounds for the decay of $B_s$ into $X_s \gamma$ and previous bounds and the green dots represents the cumulative bounds for the decay of $B$ into $\tau \nu$ and the previous bounds. All flavor observables at $95\%$ C.L.. }
    \label{chargedparticle}
\end{figure}

After applying the bounds on the plane $[ m_{h_1^+}, \, \tan \beta]$, one must project these bounds on the $[ M_{A}, \, \tan \beta]$ plane. But first, we will project the flavor bounds into the $[ M_{A}, \, m_{h_1^+}]$ plane. This is what is done in FIG. \ref{chargedparticle}. Here, we show the effect of the cumulative bounds from flavor observables in the $[ M_{A}, \, m_{h_1^+}]$ plane where the blue dots represents the allowed regions after applying the potential bounds (bounded from below conditions (BFB) and unitarity). One interesting fact is that, for high values of $m_{h_1^+}$, $M_A$ is almost degenerated with it. Then, it is impossible to have light $A$ when $h_1^+$ is a heavy particle due the conditions of the potential. This shows how important are the limits from $B \to X_s\gamma$.
For the same figure, we cumulatively added the bounds from the $B_s \to \mu \mu$ decay. These bounds are represented by red dots. It was discussed in reference \cite{Cherchiglia:2022zfy} that, for
$\tan \beta >1$, the influence of $B_s\to \mu \mu$ is marginal. However, one can observe that such statement is not exactly true, there is a non-trivial effect for simultaneous light $m_{h_1^+}$ and $M_A$.
The orange dots represents the previous cumulative bounds together with the allowed region of the inclusive radiative decay $B_s \to X_s\gamma$. Here it is clear that such bound constrain inferiorly the value of $m_{h_1^+}$, however, it still allows $M_A$ to be small. The green dots represents the previous cumulative bounds together with the allowed region of the tree-level decay $B \to \tau\nu$. Here, the bound  constrains inferiorly the value of $M_{A}$ and superiorly the value of $\tan \beta$.

\begin{figure}
    \centering
    \includegraphics[scale=0.4]{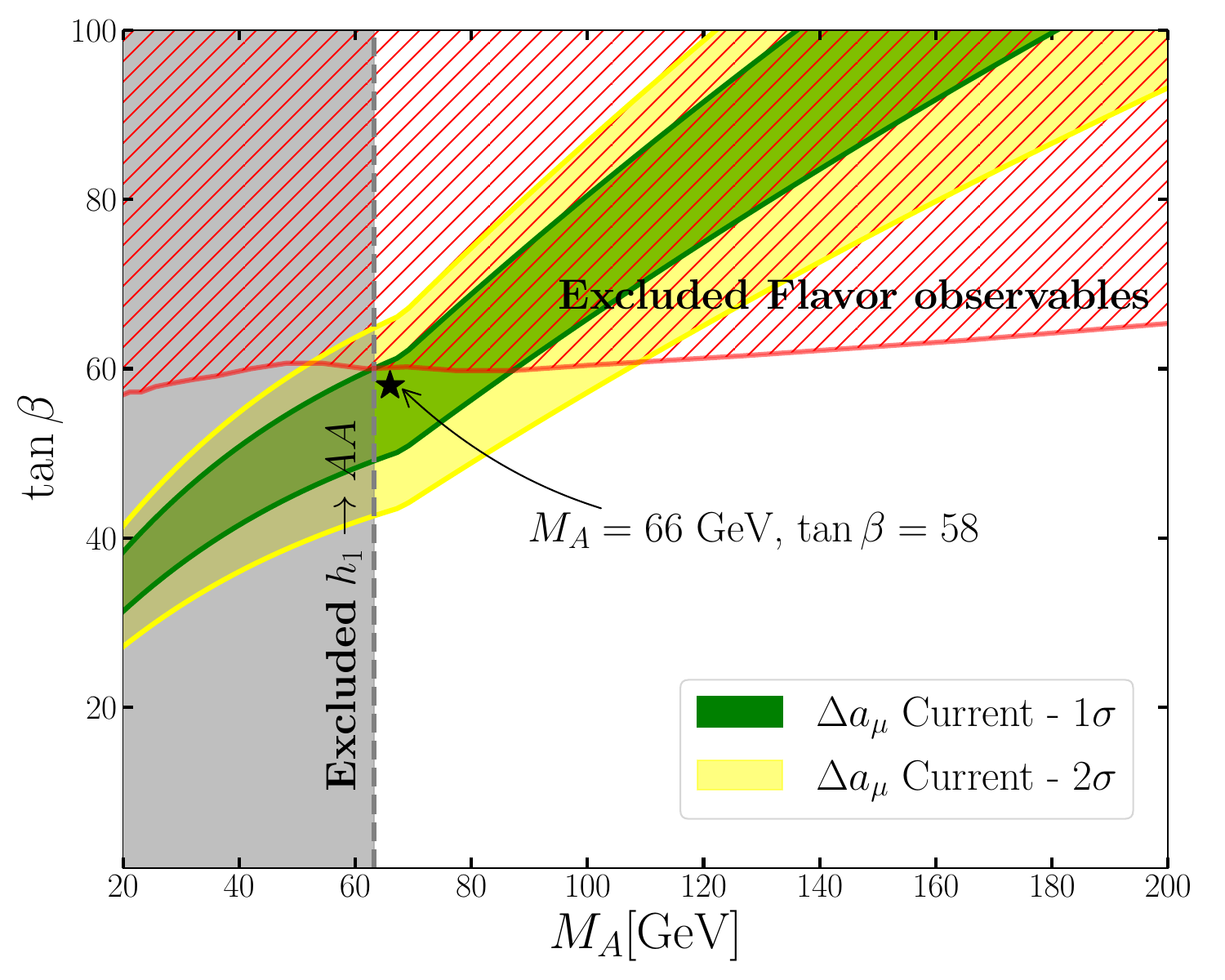}
    \caption{Imposition of $B$-physics bounds into the plane $[M_A,\tan \beta]$. The red hatched contour represents the cumulative bounds from the most stringent $B-$meson decays to limit parameter space of the pseudoscalar $A$,  $B_s \to X_s \gamma + B_s \to \mu \mu + B\to\tau\nu$ at $95\%$ C.L.. The green(yellow) band represents the current bounds from the anomalous magnetic moment of the muon for 1(2)$\sigma$. The black star represents one benchmark point that respects the $B$-physics bounds, the bounds from the invisible decay of the Higgs and explains the $\Delta a_\mu$ in the $[M_A,\tan \beta]$ plane. }
    \label{Limits_2}
\end{figure}

In FIG. \ref{Limits_2}, we projected the discussed bounds from $h_1^+$ into the $[M_A,\tan\beta]$ plane. Here, we considered the region that explains the anomalous magnetic moment of the muon for one (two) $\sigma$ as the green(yellow) contours. At the same time, as discussed in a previous section, we imposed the bounds from the invisible decay of the Higgs into two $A$'s. Such bound imposes that $M_A>62.5$ GeV, independently from the value of $\tan \beta$. After applying all discussed bounds, we discovered a small allowed region that explains the anomalous $g-2$ and at the same time obeys all main flavor constraints and do not contribute for the Higgs invisible decay.

Then, we can summarize what we did as:

1.For every point in the $[M_A,\tan\beta]$ plane, it is possible to find a parametrization for $V_L^d$ and $V_L^u$ that avoid the constraints
from neutral meson-antimeson transitions $K^0-\bar{K}^0$. Assuming
$M_A=66$ GeV and $\tan\beta=58$ as a benchmark point, we showed the
numerical values of the unitary matrices  $V_L^d$ and $V_L^u$;

2. The constraints from $B_s \to \mu \mu$ are not relevant for the
Variant I, when $\tan \beta >1$, if we simultaneous use bounds from inclusive radiative decay $b\to s \gamma$;

3. The constraint from $B \to X_s \gamma$ depend strongly if you consider FCNC. As we discussed, we found the bound:
$$m_{h^+_1}>483\,\mathrm{GeV},$$
fixing the values of $V^d_L$ and $V_L^u$ as described in the text.

4. The tree-level decay of $B\to \tau \nu$ via charged-scalar exchange
limits superiorly the value of $\tan\beta$ and gives an important bound for the plane $[M_A,\tan\beta]$. Increasing the precision of this decay is vital to rule out or not the capability of $A$ to explain the $g-2$ anomaly and run away from flavour constraints;

5. There is a small window for $(g - 2)_\mu$  be
explained in the  Variant I . The  possible solution for the
muon $g-2$ anomaly that avoids flavor constraints and higgs invisible decay lies on the range $M_A\in [62.5,122]$ GeV for $\tan \beta \in
[43,59]$;

6. In order to understand the role of $h_2$ in the calculation of  $g-2$  one needs to explore the features of the spectrum of CP-even scalars related to the  mass matrix in Eq. (5) of \cite{Pinheiro:2022bcs}. For this, we solved it numerically, scanning the parameter space as described in the beginning of this section, for tiny $f$ and $v_\eta \geq v_\rho$ while demanding $h_1$ is the $125$ GeV Higgs  of the standard model. For $\tan \beta <10$ there is no surprise.  However, when $\tan \beta > 10$ we observed a novelty exclusively for the 3-3-1 scalar potential described in \cite{Pinheiro:2022bcs}. We observed that for  $M_A>\frac{m_{h_1}}{2}$ the scan gives $M_A\approx m_{h_2}$. This is an exclusive feature of the 3-3-1. Concerning the region where  $M_A<\frac{m_{h_1}}{2}$, given the flexibility to select multiple points in the scan, we have freedom to  choose the mass of $h_2$ to be larger than the half of the Standard Higgs mass. This ensure that $h_1$ will never decay into two $h_2$ particles.

As a final comment, one benchmark point that represents the marked star in FIG. \ref{Limits_2} is: $m_{h_1}=125$ GeV, $m_{h_2}=64$ GeV, $m_{h^+}=540$ GeV, $f=2.96\times10^{-2}$ GeV, $\lambda_1=12.5,\,\lambda_2=0.85,\,\lambda_3=-6.89,\,\lambda_4=3.38,\,\lambda_5=-2.06,\,\lambda_6=-2.24,\,\lambda_9=9.34$.

\clearpage
\section{Discussion and Conclusion}
Explaining the current muon $(g-2)$ anomaly using scalar fields remains a significant challenge. At one-loop levels, charged scalars and pseudoscalars typically contribute negatively, whereas CP-even scalars contribute positively. At two-loop levels, however, all these scalars contribute positively. Individually, each scalar could potentially explain the $(g-2)$ anomaly, but they must be relatively light, which is particularly challenging for charged scalars. Neutral scalars, especially pseudoscalars, are viable candidates as they can be lighter---albeit heavier than $\frac{m_{h_1}}{2}$ to avoid invisible Higgs decays---making them natural candidates to address the $(g-2)$ anomaly.

Pseudoscalar naturally emerges in 331 model with right-handed neutrinos (331RHN). In regimes where the 331RHN features a light pseudoscalar, there exists a parameter window where a pseudoscalar mass ranging from 62.5 to 120 GeV, combined with a high $\tan \beta$, may explain the muon anomaly. We stress here that in the regime where the pseudoscalar accommodate the $g-2$ results in the CP-even $h_2$  being  degenerate in mass with the pseudoscalar $A$.  Conversely: the scenario proposed here  may be discarded if a second Higgs, heavier than the standard one, is discovered.  

Moreover, this model introduces arbitrariness regarding quark mixing, as the parametrization pattern of $V^{u,d}_L$ cannot be uniquely determined. It is plausible that a suitable parametrization of $V^{u,d}_L$ could allow heavier pseudoscalars as a solution to the $(g-2)$ anomaly, particularly if right-handed quark mixing is considered\footnote{Paper in progress.}. Thus, the choice of quark mixing pattern significantly influences all flavor physics processes involving scalars, and determining this pattern is crucial for advancing the field and achieving conclusive results in flavor physics.

In conclusion, the 331RHN model with right-handed neutrinos explain the muon $(g-2)$ anomaly through dominant two-loop contributions from a pseudoscalar with a mass in the tens of GeV range and a high $\tan \beta$, under the constraints of flavor physics. This position  the 331RHN as a compelling framework for exploring new physics beyond the Standard Model.

\section*{Acknowledgments}
We thank A. Cherchiglia for valuable discussions.
C.A.S.P  was supported by the CNPq research grants No. 311936/2021-0. J. P. P. has received support
from the European Union’s Horizon 2020 research and innovation program under the Marie Skłodowska-Curie grant agreement No 860881-HIDDeN. 
\appendix
\section{One-loop contribution}
In this work we follow the spectrum of scalars obtained in Ref. \cite{Pinheiro:2022bcs} where 
\begin{eqnarray}
    &&A=\cos \beta I_\eta +\sin \beta I_\rho,\nonumber \\
    &&h_1^+=\sin \beta \rho^+ +\cos \beta \eta^+\nonumber \\
    &&h^+_2 \approx \rho^{\prime +},
\end{eqnarray}
where $\cos \beta =\frac{v_\rho}{\sqrt{v^2_\eta + v^2_\rho}} $ and $\sin \beta =\frac{v_\eta}{\sqrt{v^2_\eta + v^2_\rho}}$.

For the case of the variant I, we have the following Yukawa interactions among $h_1^+$ and the quarks
\begin{eqnarray}
 {\cal L}&\supset& \sqrt{2} h_1^+ \left( -\frac{\cot{\beta}}{v} (V^d_L)_{ia} (V^u_L)_{bi} + \frac{\tan{\beta}}{v}(V^d_L)_{3a} (V^u_L)_{b3} \right)(m_{down})_a \bar{\hat{u}}_{b_L} \hat d_{a_R}  \nonumber \\
&+&\sqrt{2} h_1^- \left( -\frac{\tan{\beta}}{v} (V^u_L)_{ia} (V^d_L)_{bi} + \frac{\cot{\beta}}{v}(V^u_L)_{3a} (V^d_L)_{b3} \right)(m_{up})_a) \bar{\hat{d}}_{b_L} \hat u_{a_R} +H.c.
\label{Ycaseh+}
\end{eqnarray}
where the subscripts  $i=1,2$ and $a,b=1,2,3$ with $(m_{down})_{1,2,3}=(m_d\,,\,m_s\,,\,m_b)$ and $(m_{up})_{1,2,3}=(m_u\,,\,m_c\,,\,m_t)$. The fields $\hat{u}_{L,R}=(u_{L,R} \,,\, c_{L,R}\,,\, t_{L,R})$ and  $\hat{d}_{L,R}=(d_{L,R} \,,\, s_{L,R}\,,\, b_{L,R})$. Assuming the expression  above and Eqs.(\ref{MDKd}) and (\ref{MDKu}) , the effective Yukawa $y^t_{h_1^+}$ coupling of the quark top  with $h_1^+$ can be written as 
\begin{equation}
   y^t_{h_1^+} = 0.004\tan{\beta}\frac{m_t}{v} \approx 0.163 
\end{equation}
Yukawa interactions with the  higgses  $h_1$ and $h_2$
\begin{eqnarray}
 {\cal L}&\supset& \frac{\sqrt{1+\tan \beta^2}}{\tan \beta v}  \left[ c_\alpha  (V^d_L)_{ia} (V^d_L)_{bi} + s_\alpha \tan{\beta}(V^d_L)_{3a} (V^d_L)_{b3} \right](m_{down})_a \bar{\hat{d}}_{b_L} \hat d_{a_R}h_1  \nonumber \\
&+&\frac{\sqrt{1+\tan \beta^2}}{\tan \beta v}  \left[ s_\alpha \tan{\beta} (V^u_L)_{ia} (V^u_L)_{bi} + c_\alpha (V^u_L)_{3a} (V^u_L)_{b3} \right](m_{up})_a \bar{\hat{u}}_{b_L} \hat u_{a_R}h_1 +H.c,
\label{Ycaseh1}
\end{eqnarray}
and
\begin{eqnarray}
 {\cal L}&\supset& \frac{\sqrt{1+\tan \beta^2}}{\tan \beta v}  \left[ -s_\alpha  (V^d_L)_{ia} (V^d_L)_{bi} + c_\alpha \tan{\beta}(V^d_L)_{3a} (V^d_L)_{b3} \right](m_{down})_a \bar{\hat{d}}_{b_L} \hat d_{a_R}h_2 \nonumber \\
 &+&\frac{\sqrt{1+\tan \beta^2}}{\tan \beta v}  \left[ -c_\alpha \tan{\beta} (V^u_L)_{ia} (V^u_L)_{bi} - s_\alpha (V^u_L)_{3a} (V^u_L)_{b3} \right](m_{up})_a \bar{\hat{u}}_{b_L} \hat u_{a_R}h_2+H.c.
\label{Ycaseh2}
\end{eqnarray}
where $s_\alpha=\sin{\alpha}$ and $c_\alpha=\cos{\alpha}$ are the mixing angle among the neutral higgses $R_\eta$ and $R_\rho$ tht compose $h_1$ and $h_2$ following Ref. \cite{Pinheiro:2022bcs}

The Yukawa interactions with  $A$ is given by
\begin{eqnarray}
 {\cal L}_Y^{A}&=&iA  \bar{\hat{u}}_{b_L}\left(-\frac{\tan{\beta}}{v} (V^u_L)_{ia} (V^u_L)_{bi}(m_{up})_a+ \frac{ \cot{\beta} }{v} (V^u_L)_{3a} (V^u_L)_{b3}(m_{up})_a    \right)\hat u_{a_R} \nonumber \\
&+& iA \bar{ \hat{d}}_{b_L}\left( \frac{\cot{\beta}}{v} (V^d_L)_{ia} (V^d_L)_{bi}(m_{down})_a   +\frac{\tan{\beta}}{v} (V^d_L)_{3a} (V^d_L)_{b3}(m_{down})_a\right)\hat d_{a_R}  + \text{H.c.}\,,
\label{YcaseIA}
\end{eqnarray}
\noindent the indexes $i=1,2$ and $a,b=1,2,3$  were defined in the previous expressions. 
Considering  the expression above and Eqs.(\ref{MDKd}) and (\ref{MDKu}) , the effective Yukawa $y^t_{A}$ coupling of the  quark top with $A$ can be written as 
\begin{equation}
   y^t_{A} = 0.004\tan{\beta}\frac{m_t}{v}  \approx 0.163. 
\end{equation}
\section{One and two-loop contribution of CP-odd scalar}

The light CP-odd scalars negative one-loop contribution for the anomalous magnetic moment of the muon is given by:
\begin{eqnarray}
    \Delta a_\mu(A)^\mathrm{(1-loop)}=-  \frac{m_\mu^2}{8\pi^2M_A^2 }\Big( \frac{g^2m_\mu^2A_\mu^2}{4M_W^2} \Big) H(\frac{m_\mu^2}{M_A^2}),
    \label{oneloop}
\end{eqnarray}
such that $H(y)=\int_0^1\frac{x^3dx}{1-x+x^2y}$.
The light CP-odd scalars positive two-loop contribution for the anomalous magnetic moment of the muon is given by:

\begin{eqnarray}
    \Delta a_\mu(A)^\mathrm{(2-loop)}=\Delta a_\mu(A)^\mathrm{(2-loop)}_{\gamma} + \Delta a_\mu(A)^\mathrm{(2-loop)}_{Z},
\end{eqnarray}
such that:
\begin{eqnarray}
    \Delta a_\mu(A)^\mathrm{(2-loop)}_{\gamma}=  \frac{\alpha^2}{8\pi^2\sin^2\theta_W }\frac{m_\mu^2A_\mu}{M_W^2}\sum_{f=t,b,\tau} N_c^f q_f^2 A_f \frac{m_f^2}{M_A^2}  {\cal F}\Bigg(\frac{ m_f^2}{M_A^2}\Bigg)
    \label{twoloop_1}
\end{eqnarray}
and

\begin{eqnarray}
    \Delta a_\mu(A)^\mathrm{(2-loop)}_{Z}=  \frac{\alpha^2 m_\mu^2 A_\mu g_V^\mu}{8\pi^2\sin^4\theta_W \cos^4\theta_W M_Z^2}\sum_{f=t,b,\tau}  \frac{N_c^f g_V^f q_f A_f m_f^2}{M_Z^2-M_A^2}  \Bigg[ {\cal F}\Bigg(\frac{ m_f^2}{M_Z^2}\Bigg)-{\cal F}\Bigg(\frac{ m_f^2}{M_A^2}\Bigg) \Bigg]
\end{eqnarray}
such that $g_V^f =\frac{1}{2}T_3(f_L)-q_f\sin^2\theta_W $ and ${\cal F}(x)=\int_0^1 dz \ln \Big( \frac{x}{z(1-z)}\Big)\frac{1}{x-z(1-z)}$.

 We performed the full calculation using the analytical formulas from \cite{Chun:2019oix}, adding charged and CP-even scalars too. 

\section{Asymptotic Limits}

Analyzing  FIG. \ref{fig:assimptoticlimit}, it is possible to observe, for sufficiently high values of $\tan\beta$, a particular value of $M_A$ in which the 2-loop contribution for the $g-2$ of the muon becomes dominant in relation to the 1-loop one.

Here in this appendix, we will try to estimate analytically  which value of $M_A$  the two-loop contribution becomes dominant. One first observation is that  $\Delta a_\mu(A)^\mathrm{(2-loop)}_{Z}$ is always suppressed in relation to $\Delta a_\mu(A)^\mathrm{(2-loop)}_{\gamma}$. Then, it is sufficient to understand the limit in which the photon Barr-Zee contribution is larger than the negative one-loop contribution, or,  $\Delta a_\mu(A)^\mathrm{(2-loop)}_{\gamma}  + \Delta a_\mu(A)^\mathrm{(1-loop)}=0$.

Using equations \ref{oneloop} and \ref{twoloop_1}, we will obtain the approximated equation:

\begin{eqnarray}
    \frac{1}{3}m_b^2 {\cal F}\Bigg( \frac{m_b^2}{M_A}\Bigg) + \frac{4}{3 \tan^4 \beta}m_t^2 \ln\Bigg(\frac{m_t^2}{M_A^2}\Bigg) + m_\tau^2 {\cal F}\Bigg( \frac{m_\tau^2}{M_A}\Bigg) = \frac{g^2 s_W^2 m_\mu^2}{4\alpha^2} H\Bigg(\frac{m_\mu^2}{M_A^2} \Bigg)
\label{asymptotic}
\end{eqnarray}

For $\tan \beta$ larger than one and smaller than 5, the top-quark contribution is relevant to the total magnetic moment of the muon. However, for such small values of $\tan \beta$, it is not possible for $A$ to explain the anomaly(FIG. \ref{fig:assimptoticlimit}). For $\tan \beta > 5$, the top-quark contribution is irrelevant to the total magnetic moment of the muon. Not only that, $\Delta a_\mu(A)^\mathrm{(2-loop)}_{\gamma} = - \Delta a_\mu(A)^\mathrm{(1-loop)}$ becomes independent of $\tan \beta$, as observed in FIG. \ref{fig:assimptoticlimit}, too.

Now, solving numerically:
\begin{eqnarray}
    \frac{1}{3}m_b^2 {\cal F}\Bigg( \frac{m_b^2}{M_A}\Bigg) + m_\tau^2 {\cal F}\Bigg( \frac{m_\tau^2}{M_A^2}\Bigg) = \frac{g^2 s_W^2 m_\mu^2}{4\alpha^2} H\Bigg(\frac{m_\mu^2}{M_A^2} \Bigg)
\label{asymptotic_2}
\end{eqnarray}
leads to $M_A \approx 3.16$ GeV, explaining the shape of the figures in FIG. \ref{fig:assimptoticlimit}.

\bibliography{references} 

\end{document}